\documentclass[twocolumn,prl,aps,superscriptaddress,showpacs,amsmath,amssymb,floatfix]{revtex4-1}
\usepackage{graphicx}
\usepackage{amssymb}
\usepackage{amsmath}
\usepackage{epsfig}
\usepackage{color}
\usepackage{mathtools}
\usepackage[colorlinks,linkcolor=blue,anchorcolor=blue,citecolor=blue,urlcolor=blue]{hyperref}
\usepackage{physics}
\usepackage{bm}
\begin{document}

\title{Topological indexes in symmetry preserving dynamics}
\author{Ze-Gang Liu}
\affiliation{CAS Key Laboratory of Quantum Information, University of Science and Technology of China, Hefei, 230026, People’s Republic of China}
\author{Long Xiong}
\affiliation{CAS Key Laboratory of Quantum Information, University of Science and Technology of China, Hefei, 230026, People’s Republic of China}
\author{Beibing Huang}
\affiliation{Department of Physics, Yancheng Institute of Technology, Yancheng, 224051, China}
\author{Guang-Can Guo}
\affiliation{CAS Key Laboratory of Quantum Information, University of Science and Technology of China, Hefei, 230026, People’s Republic of China}
\affiliation{Synergetic Innovation Center of Quantum Information and Quantum Physics, University of Science and Technology of China, Hefei, 230026, P.R. China}
\author{Ming Gong}
\email{gongm@ustc.edu.cn}
\affiliation{CAS Key Laboratory of Quantum Information, University of Science and Technology of China, Hefei, 230026, People’s Republic of China}
\affiliation{Synergetic Innovation Center of Quantum Information and Quantum Physics, University of Science and Technology of China, Hefei, 230026, P.R. China}
\date{\today }

\begin{abstract}
The quench dynamics of topological phases have received intensive investigations in recent years. In this work, we prove exactly that 
the topological invariants for both $\mathbb{Z}$ and $\mathbb{Z}_2$ indexes are independent of time in symmetry preserving dynamics. 
We first reach this conclusion by a direct relation between the time derivative of Berry connection and the Hamiltonian energy based on the time dependent 
Hellman-Feynman theorem, with which we show exactly that the topological indexes for systems without and with time reversal symmetry are unchanged during evolution.
In contrast, the geometry phase without symmetry protection in a closed parameter space can change dramtically with time, as revealed from the parameterized Landau-Zener model. Then we interpret this result by showing that the time dependent wave function is essentially the eigenvector of an auxiliary Hamiltonian, which 
has exactly the same spectra and symmetries as the original Hamiltonian. For this reason, the adiabatic evolution between the original and auxiliary Hamiltonian will not lead to gap closing and reopening, thus the topological indexes are independent of time. This result has generality and can be applied to models with other symmetries and
dimensions, and may even be applied to gapless phases. Finally, possible ways to outreach this rigorous result are discussed.
\end{abstract}
\maketitle

Topological phases have become one of the most thriving research fields in recent years. These topological phases are characterized by integer indexes and 
support topological protected edge modes at the boundaries and defects. In these phases, the topological insulators, which may 
have important applications in spintronics and other quantum devices\cite{qi2011topological,chen2009experimental,chang2013experimental,zhang2010crossover}, have already 
been realized in a lot of solid materials; see a recent review in Ref. \cite{ando2013topological}. These materials are protected by time reversal symmetry (TRS), and are 
characterized by $\mathbb{Z}_2$ index in two dimensions and $\mathbb{Z}$ index in three dimensions\cite{schnyder2008classification, ryu2010topological, 
kitaev2009periodic}. Topological superconducting phases protected by the intrinsic particle-hole symmetry (PHS) in the Bogoliubov-de Gennes formulism\cite{de1989superconductivity}
have also been widely investigated. This system has the potential to realize Majorana zero modes at the ends or vortex cores\cite{kitaev2001unpaired, read2000paired,ivanov2001non}. 
The topological indexes, similar to topological insulators, also depend strongly on the dimensions and symmetries. In recent years, several experimental groups have 
reported strong evidences for existence of Majorana zero modes in spin-orbit coupled systems in proximity to $s$-wave superconductors\cite{mourik2012signatures, das2012zero, nadj2014observation, deng2012anomalous, sun2016majorana, 
he2017chiral}. In ultracold atoms, the spin-orbit coupling, which is the key ingredient in topological phases, have also been realized in alkali and rare-earth atoms using 
different methods\cite{aidelsburger2014measuring,jotzu2014experimental,galitski2013spin, zhang2013topological,wu2016realization}. These achievements shed great light on the realization of topological quantum computation with these protected excitations.

With these progresses, it is quite natural to explore the dynamics of topological phases. These systems provide an excellent opportunity to explore the 
dynamics of discrete values under continuous evolution. In Ref. \cite{dong2015dynamical}, Dong {\it et al} show that the  Chern number is not 
changed during time evolution process in a four-band spin-orbit coupled topological superfluids. More rigorous conclusions are presented in Refs. \cite{d2015dynamical,wang2017scheme,caio2015quantum}
based on two-band models. In Ref. \cite{d2015dynamical}, the dynamics of the wave function is mapped to a vector rotating around an effective magnetic field in momentum space, and the Skyrmion number is proved exactly to be time independent. 
In Ref. \cite{caio2015quantum}, the non-equilibrium response in a two-band model is calculated, which is also independent of time. However, the edge modes and corresponding currents may carry information about the topological phase transitions. In Ref. \cite{wang2017scheme}, Wang {\it et al} even point out the intimate relation between  linking number and the topological indexes of the quenched Hamiltonian, which has recently been experimentally verified\cite{tarnowski2017characterizing}. 
All these results have a common feature, that is, all models are considered in two spatial dimensions without 
TRS. However, since the topological indexes are impossible to be changed continuously, this conclusion may be quite general, that is, it may be generalized to models with other symmetries and dimensions\cite{schnyder2008classification, ryu2010topological, kitaev2009periodic}, which is the major motivation of this work. 
In regarding of the unprecedented experimental flexibility in ultracold atoms in the realization of different physical models\cite{flaschner2016experimental, flaschner2016observation,sadler2006spontaneous,chang2004observation,schmaljohann2004dynamics,chang2005coherent,hendry1994generation,
chuang1991cosmology,bauerle1996laboratory,aidelsburger2014measuring,jotzu2014experimental,galitski2013spin,zhang2013topological,wu2016realization}, including the 
topological models\cite{aidelsburger2014measuring,jotzu2014experimental,galitski2013spin,zhang2013topological,wu2016realization} and the dynamics of many-body states\cite{kinoshita2006quantum,
schreiber2015observation, baumann2010dicke,deng2016observation, kaufman2016quantum}, the dynamics of topological phases should be able to reveal some more intriguing features that not seen in topological trivial phases.

In this work, we explore the dynamics of topological indexes during symmetry preserving process in some more general models with and without TRS.
We show that these indexes are independent of time. 
This conclusion is related to that in quantum Hall states\cite{hatsugai1993chern,thouless1982quantized,paalanen1982quantized,von1986quantized,zhang2005experimental} and
topological superconducting phases\cite{read2000paired, qi2010chiral,leijnse2012introduction} without TRS, which are characterized by $\mathbb{Z}$ index; and in systems with TRS such as 
topological insulators\cite{qi2006topological,fu2007topological,fu2007topological,bernevig2006quantum,qi2008topological,qi2010quantum}, which are characterized by $\mathbb{Z}_2$ index\cite{fu2006time}. 
We establish this generic conclusion using the time dependent Hellman-Feynman theorem\cite{feynman1939forces,epstein1966time, di2000hellmann}, in which the time derivative of Berry connection is directly related to change of Hamiltonian energy. We reach this conclusion by direct calculation of the topological indexes. In contrast, the geometry phase in a closed loop in parameter space may change dramatically with time, which is demonstrated by a parameterized Landau-Zener model. Next, we show that the time dependent wave function is the eigenvector of an auxiliary Hamiltonian, which has the same spectra and symmetries as the original Hamiltonian. For this reason, the
adiabatic evolution between the auxiliary Hamiltonian and the original Hamiltonian will not involve gap closing and reopening, thus the topological indexes are independent of time. The latter picture naturally extends our conclusion to more general models with different dimensions, symmetries and topological indexes, and even for topological gapless 
phases. Finally, we discuss the applicability of our result and the possible ways to outreach the rigorous result.

We start from the following generalized Hellman-Feynman theorem\cite{feynman1939forces,epstein1966time, di2000hellmann}. For the time dependent Schr\"odinger equation ($\lambda$ serves as parameter space),
\begin{equation}
    i \partial_t |\psi(\lambda, t)\rangle = H(\lambda, t) |\psi(\lambda, t)\rangle,
\end{equation}
the Berry connection with respect to parameter $\lambda$ at time $t$ is defined as
\begin{equation}
    \mathcal{A}(\lambda,t) = i \langle \psi(\lambda,t) | \partial_\lambda | \psi(\lambda,t)\rangle.
\end{equation}
The dynamics of this connection is defined as,
\begin{equation}
    \dot{\mathcal{A}}(\lambda,t) = i (\langle \dot{\psi}(\lambda,t) | \partial_\lambda | \psi(\lambda,t)\rangle + \langle \psi(\lambda,t) | \partial_\lambda | 
	\dot{\psi}(\lambda,t)\rangle).
\end{equation}
Inserting the Schr\"odinger equation into the above equation, we immediately obtain the following key result,
\begin{equation}
    \dot{\mathcal{A}}(\lambda, t) = \langle \psi(\lambda, t) | ({\partial H \over \partial \lambda}) | \psi(\lambda, t)\rangle.
    \label{eq-gH}
\end{equation}
We can recover the static Hellman-Feynman theorem for a time independent Hamiltonian by setting $|\psi(\lambda, t)\rangle = |\psi(\lambda)\rangle e^{-iE_\lambda t}$, with which the Berry connection is 
reduced to $\mathcal{A}(\lambda, t) = {\partial E_\lambda \over \partial \lambda} t + i \langle \psi(\lambda) | \partial_\lambda| \psi(\lambda) \rangle  $, and the right hand side of Eq.\ (\ref{eq-gH}) 
becomes $\dot{\mathcal{A}}(\lambda, t)  = {\partial E_\lambda \over \partial \lambda}$. We define the right band side of Eq. \ref{eq-gH} as Hamiltonian energy throughout this work, since it is related to the change of energy of the Hamiltonian. This quantity is gauge invariant for time independent gauge potential. In following, we denote this energy as $\epsilon_\lambda$ for convenient, that is, 
$\dot{\mathcal{A}}(\lambda, t) = \epsilon_\lambda$. Obviously, in the adiabatic limit, 
$\epsilon_\lambda = {\partial E_\lambda \over \partial \lambda}$, the dynamics of geometry phase along a closed loop follows exactly,
\begin{eqnarray}
	\dot{\gamma}(t) = \oint \epsilon_\lambda d \lambda = \oint {\partial E_\lambda \over \partial \lambda} d \lambda= 0.
\label{eq-gammadot}
\end{eqnarray}

{\it Dynamics of geometry phase $\gamma$}. We first present some exact results for the dynamics of topological indexes, which reveals some 
beautiful mathematical features concealed in the topological indexes. Then, we will try to understand the exactly results in a more generic way, from which these exact results can be extended to 
more general models with other dimensions and symmetries. Let us first consider the dynamics of geometry phase, which has geometric interpretation, but does not protected by any symmetry. Thus the geometry phase can take any value in a close loop in parameter space. We consider the following 
parameterized Landau-Zener model,
\begin{equation}
H = v t \sigma_z + g(\sigma_x \cos{\lambda} - \sigma_y \sin{\lambda}).
\end{equation}
The wave function for $\lambda \in [0, 2 \pi)$ can be written as,
\begin{equation}
    |\psi(\lambda, t)\rangle = (a, be^{-i\lambda})^T,
\end{equation}
where $(a, b)^T$ is the solution of the standard Landau-Zener model\cite{zener1932non, wittig2005landau, shevchenko2010landau}, $\mathcal{H}_\text{LZ} = vt \sigma_z + 
g \sigma_x$, with,
\begin{equation}
    \gamma(t) = 2\pi |b(t)|^2,
\end{equation}
thus the geometry phase should be time dependent. With the exact expression in the Landau-Zener model for $|\psi(t \rightarrow -\infty)\rangle = (1,0)^T$, we 
have $\lim_{t\rightarrow +\infty}\gamma(t) = 2\pi(1- e^{-\pi g^2/ v})$, which reduces to Eq. \ref{eq-gammadot} in the adiabatic limit with $v\rightarrow 0$. However, when $t\sim 0$, the coupling between the two bands can lead to strong oscillation for geometric phase $\gamma(t)$ and its time derivative $\gamma'(t) = 4\pi g \Im (b^\ast a)$. 

This is different from the condition when the geometric phase is quantized by introducing some extra symmetries to the Hamiltonian, in which case $\gamma(t) = 2\pi n$, $n \in \mathbb{Z}$.
The evolution of this number can be rewritten as
\begin{equation}
    \gamma(t+\delta t) - \gamma(t) = (\oint \epsilon_\lambda d \lambda) \delta t.
    \label{eq-discrete}
\end{equation}
Here we assume $\epsilon_\lambda \ne {\partial E_\lambda \over \partial \lambda}$. Notice that the right hand side is always a continuous  function of $t$ for a smooth quench process. Let $\text{supp} (\oint |\epsilon_\lambda| d\lambda) = \mathcal{M} < \infty$, then we may always find a sufficient small $\delta t$ to ensure that $|\gamma(t+\delta t) - \gamma(t)| \le \mathcal{M} \delta t < 2\pi$, for which reason discrete jump of the geometric phase between two discrete values is not allowed. This result presents an intuitive picture to understand the dynamics of topological invariants during continuous evolution. More generic way to this result will be presented below.

\begin{figure}
    \centering
   \includegraphics[width=0.45\textwidth] {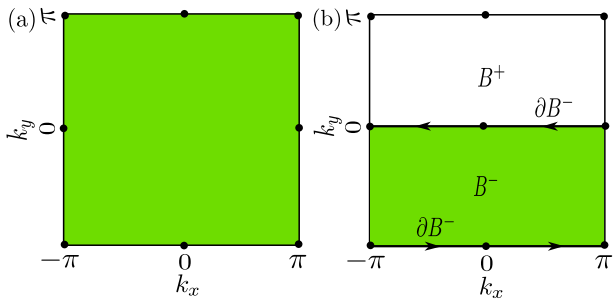}
   \caption{(Color online) BZ of the square lattice without TRS (a), and with TRS (b). In (b) the BZ is 
    divided into two half BZs, $B^+$ and $B^-$, and the thick lines with arrows mark the boundary of $B^-$, denoted as $\partial B^-$. 
	The green regimes are used to computer the topological indexes.}
    \label{fig-fig1}
\end{figure}

{\it Dynamics of $\mathbb{Z}$ index}. We turn to the dynamics of Chern number in two spatial dimensions for systems without TRS. The typical candidates include integer  
quantum Hall states, anomalous quantum Hall states\cite{hatsugai1993chern,thouless1982quantized,paalanen1982quantized,von1986quantized,zhang2005experimental}, and topological superconducting 
phases\cite{read2000paired,qi2010chiral,leijnse2012introduction}. In following, we consider the physics in an square lattice with lattice constant $a=1$. We set the parameter $\lambda$ to ${\bm k} = (k_x, k_y)$, and drop the band index $n$ for simplicity. Our consideration is that, since the index is invariant in each band, it should be invariant 
in the whole system after summation of contribution from all occupied bands. The typical Brillouin zone (BZ) is presented in Fig. \ref{fig-fig1}a, and the Chern number in this
closed torus is defined as,
\begin{eqnarray}
C(t)= \int_{-\pi}^{\pi} d k_x \int_{-\pi}^\pi d k_y  \bigtriangledown_k \times \mathcal{A}({\bm k}, t)_z.
\end{eqnarray}
Using the time dependent Hellman-Feynman theorem in Eq. \ref{eq-gH}, we find 
\begin{eqnarray}
    \partial_t C  \nonumber 
    &=&  \int  ( \partial_{k_x} \dot{\mathcal{A}}_y - \partial_{k_y} \dot{\mathcal{A}}_x ) d^2{\bm k}=\int \left( \partial_{k_x} \epsilon_y - \partial_{k_y} \epsilon_x \right) d^2{\bm k} \nonumber \\
&=& \int_{-\pi}^{\pi}  \left[ \epsilon_y(\pi,k_y,t)-\epsilon_y(-\pi,k_y,t) \right] d k_y -  \nonumber \\
& & \int_{-\pi}^{\pi}  \left[ \epsilon_x(k_x,\pi,t)-\epsilon_x(k_x,-\pi,t) \right] d k_x =0.
    \label{eq-Z0}
\end{eqnarray}
In above equation, $\epsilon_x(k_x,\pi,t)=\epsilon_x(k_x,-\pi,t)$ since $|\psi_n(k_x,\pi,t)\rangle$ and $|\psi_n(k_x,-\pi,t)\rangle$ are the same physical state due to the periodicity of the reciprocal momentum space, which can at most differ by a global phase, {\it i.e.}, $|\psi_n(k_x,\pi,t)\rangle =\exp[i \theta_x(k_x)] |\psi_n(k_x,-\pi,t)\rangle$. This phase will not enter the Hamiltonian 
energy due to its gauge invariant in Eq. \ref{eq-gH}. The same analysis holds for $\epsilon_y(\pi,k_y,t) = \epsilon_y(-\pi,k_y,t)$, thus the above result should be exactly equal to zero. Notice 
that the above analysis is quite similar to the analysis of the integer property of Chern number\cite{shen2012topological, kohmoto1985topological}. The only difference is that for Berry connection, 
the periodicity in momentum space ensures that the Berry connection may differ by $2\pi n$, where $n \in \mathbb{Z}$, between the two equivalent points. In our model only $n = 0$ is allowed.
Obviously, this conclusion holds for arbitrary number of occupied bands, thus the Chern number for all occupied bands is independent of time. In this way, we generalize the conclusion in Refs. \onlinecite{dong2015dynamical,d2015dynamical, caio2015quantum} to arbitrary bands. 

{\it Dynamics of $\mathbb{Z}_2$ index}. The topological Chern number is defined in systems without any symmetry. However, symmetries, such as TRS, PHS 
and chiral symmetries {\it etc}, are important and essential in the characterization of different topological phases. Here we mainly focus on the TRS, in which the topological index is reduced from $\mathbb{Z}$
to $\mathbb{Z}_2$. In the time evolution process, this index is well-defined {\it only} when the dynamical wave functions still respect TRS, that is,
\begin{equation}
T\psi_{{\bm k}}^\sigma(t) = \psi_{-{\bm k}}^{\bar{\sigma}}(t),
    \label{eq-TO}
\end{equation}
where $\sigma = \uparrow$ ($\downarrow$), $\bar{\sigma} = \downarrow$ ($\uparrow$). In above, the TRS operator is defined as $T = u_T \mathcal{K}$, where $\mathcal{K}$ is the complex conjugate and $u_T$ is an 
unitary operator, {\it i.e.},  $u_T u_T^\dagger = 1$.  Let us define $\psi_{{\bm k}}^\sigma(t) = \mathcal{U}_{\bm k}(t) \psi_{\bm k}^\sigma(0)$, where $\mathcal{U}_{{\bm k}}(t) = \prod_{i=N}^1 
e^{-iH_{{\bm k}}(t_i) \delta t}$. Eq. \ref{eq-TO} requires that $u_T H_{\bm k}^\ast(t) u_T^\dagger = -H_{-{\bm k}}(t)$, which leads to
\begin{equation}
    T \mathcal{U}_{\bm k}(t) T^{-1} = \mathcal{U}_{-{\bm k}}(t).
\end{equation}
In this way, the quenched Hamiltonian no longer possesses TRS. This constraint is similar, but different from the PHS, since $u_T \ne \sigma_x$. Notice that our consideration here is different from that in previous literatures, where the TRS can only be well defined in one full modulating period. For instance, in Refs. \cite{carpentier2015topological,yao2017topological,yan2015general}, the topological $\mathbb{Z}_2$ in periodically modulating systems is constructed in different ways to recover the TRS. The similar consideration should be taken into account seriously for models with chiral symmetry. Our definition of $H_{\bm k}(t)$ ensures the definition of $\mathbb{Z}_2$ index in any quench process, not limited to periodic modulation. 

We understand the invariant of this index from two different levels. The minimal model with $\mathbb{Z}_2$ index should be at least a four-band model, such as the Bernevig-Hughes-Zhang model\cite{bernevig2006quantum,zhou2008finite}, which can be written as
\begin{equation}
    H_\text{BHZ}({\bm k}) = \begin{pmatrix}
        h_\uparrow({\bm k})  &  0 \\
        0           & h_\downarrow({\bm k})
    \end{pmatrix}.
\end{equation}
The whole system still respect the TRS, although each block does not. In each block we can define the corresponding Chern number $C_\sigma(t)$ for $\sigma=\uparrow, \downarrow$. 
The TRS ensures that $C_\uparrow(t) + C_\downarrow(t) = 0$ due to the properties of wave function in Eq. \ref{eq-TO}. The $\mathbb{Z}_2$ index is defined as 
$C_2 = (C_\uparrow(t) - C_\downarrow(t))/2$ mod(2)\cite{fu2006time}, thus $C_2(t)$ is independent of time from Eq. \ref{eq-Z0}.

With this result in mind, we next move the more general models without the above special feature. In this case, the topological index is defined as \cite{fu2006time, fukui2007topological},
\begin{eqnarray}
C_2 &=& \dfrac{1}{2\pi i} \left[\oint_{\partial B^-} d \bm{k} \cdot \mathcal{A} - \int_{B^-} d^2 k \mathcal{F} \right],
\label{Z2}
\end{eqnarray}
where the integration is performed over a half BZ, $B^-=\left[-\pi,\pi \right] \otimes \left[ -\pi, 0 \right]$, and its boundary $\partial B^-$; 
see Fig.\ \ref{fig-fig1}b. Eq. \ref{eq-TO} ensures that the two wave functions with opposite momenta should have opposite Berry connection, thus the integration of Berry culvature $\mathcal{F}$ over the whole 
BZ should be exactly equal to zero, similar to that in the static Hamiltonian. In above equation, the first term counteract the effect of boundary in the half BZ. With the similar techniques in Eq. \ref{eq-Z0}, we find 
\begin{eqnarray}
\partial_t \oint_{\partial B^-} d \bm{k} \cdot \mathcal{A} =\int_{-\pi}^{\pi} \left[ \epsilon_x(k_x,-\pi) -\epsilon_x(k_x,0) \right] d k_x,
\label{Z2_A}
\end{eqnarray}
and 
\begin{eqnarray}
\partial_t \int_{B^-} d^2{\bm k} \mathcal{F} && = \int_{-\pi}^{0}  \left[ \epsilon_y(\pi,k_y) - \epsilon_y(-\pi,k_y) \right] d k_y \nonumber \\ 
&&- \int_{-\pi}^{\pi}  \left[ \epsilon_x(k_x,0) - \epsilon_x(k_x,-\pi) \right] d k_x,
    \label{Z2_F}
\end{eqnarray}
The periodicity in Hamiltonian energy in $k_x$ direction means $\epsilon_y(\pi,k_y) - \epsilon_y(-\pi,k_y) = 0$. The other two terms with contribution from the bulk and boundary  are exactly cancelled from Eqs. \ref{Z2_A} and \ref{Z2_F}, we immediately find,
\begin{equation}
    \partial_t C_2 = 0. 
    \label{eq-C2t}
\end{equation}
This proof is independent of the number of bands involved, thus this conclusion is also valid for arbitrary number of bands.

\begin{figure}
	\centering
	\includegraphics[width=0.36\textwidth]{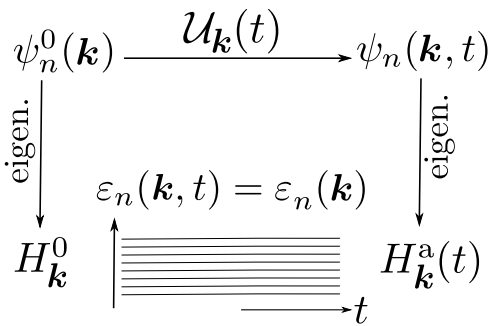}
	\caption{(Color online). Equivalent principle for the dynamics of topological indexes. The wave function $|\psi_{n}({\bm k},t)\rangle$ can 
        be regarded as the eigenvector (eigen.) of an auxiliary Hamiltonian $H_{\bm k}^\text{a}$ with  the same spectra and symmetries as $H_{\bm k}^0$. }
	\label{fig-fig2}
\end{figure}

{\it Interpretation of this result and generalization to other dimensions and symmetries}. The above proofs reveal the beautiful mathematical features in the 
topological indexes. These structures can be used to study the dynamics of other topological invariants, such as Hopf links and knots\cite{satija2017observation, deng2016probe}, linking numbers\cite{wang2017scheme} and higher dimensional non-Abelian 
topological indexes\cite{wang2012simplified}, which are also independent of time. Eq. \ref{eq-discrete} provides an intuitive picture why continuous change can not lead to discrete jumps. Now, we try to seek a more fundamental interpretation of this result. Let $|\psi_n^0({\bm k})\rangle$ being the eigenvector of
$H_{\bm k}^0$, that is, $H_{\bm k}^0 |\psi_n^0({\bm k})\rangle = \varepsilon_n({\bm k}) |\psi_n^0({\bm k})\rangle$. We can define the wave function at any time $t$ via the 
transformation $|\psi_n({\bm k}, t) \rangle = \mathcal{U}_{\bm k}(t) |\psi_n^0({\bm k})\rangle$. This new instantaneous wave function can be regarded as the eigenvector of the following 
auxiliary Hamiltonian, 
\begin{equation}
    H_{\bm k}^\text{a}(t) |\psi_n({\bm k}, t)\rangle = \varepsilon_n({\bm k}) |\psi_n({\bm k}, t)\rangle,
    \label{eq-lambdai}
\end{equation} 
where $H_{\bm k}^\text{a} = \mathcal{U}_{\bm k}(t) H_{\bm k}^0 \mathcal{U}_{\bm k}^\dagger(t)$. The auxiliary Hamiltonian has the same symmetry as the origin Hamiltonian.
For instance, when $T_0$ is the TRS operator for the initial Hamiltonian $H_{\bm k}^0$, then the corresponding TRS for the auxiliary Hamiltonian should be $T_a = \mathcal{U}_{-{\bm k}} T_0 \mathcal{U}_{\bm k}^\dagger = \mathcal{U}_{-{\bm k}} u_0 \mathcal{U}^T_{{\bm k}} \mathcal{K}$, where $T_0 = u_0 \mathcal{K}$.
This result indicate that asking the topological invariant of the instantaneous wave function at time $t$ is equivalent to ask the question of topological 
invariant of the auxiliary Hamiltonian (see Fig, \ref{fig-fig2}). Since the band structures and symmetries for $H_{\bm k}^0$ and $H_{\bm k}^\text{a}$ are exactly the same, the adiabatic transition from the original
Hamiltonian and the new Hamiltonian does not involve the gap closing and reopening, thus the topological index should be independent of time. 
The same analysis can be applied to study the topological invariants with chiral symmetry. This proof is rather general, and can be applied to understand the topological 
indexes in other dimensions. Our conclusion may also be true for gapless phases, and even for the many-body systems in which some of the parameters need to be solved 
self-consistently\cite{dong2015dynamical}. 

In these two Hamiltonians, the relation between the Berry connection can be written as,
\begin{equation}
    \mathcal{A}({\bm k},t) = \mathcal{A}({\bm k}) +i \langle \psi_n^0({\bm k})|\mathcal{U}^\dagger_{\bm k}(t) (\partial_{\bm k} \mathcal{U}_{\bm k}(t))| \psi_n^0({\bm k})\rangle,
\end{equation}
where $\mathcal{A}({\bm k})$ and $\mathcal{A}({\bm k},t)$ are the initial and final Berry connections. Take the time derivative to the above 
expression will direct yields Eq. \ref{eq-gH}. The general conclusion in this work means that $\mathcal{A}({\bm k}, t)$ can be smoothly connected to 
$\mathcal{A}({\bm k})$ since they belong to the same phase. Let $\mathcal{U}_{\bm k}= \exp(i\theta_{\bm k})$, then $\mathcal{A}({\bm k},t) = \mathcal{A}({\bm k})- \partial_{\bm k} \theta_{\bm k}$,
which is just the Abelian gauge potential. Although this term will not change the topological index, it will change the Berry curvature and the associated semiclassical trajectories\cite{xiao2010berry, chang1996berry}. 

{\it Discussion and remarks}. Finally, several remarks about these results in this work are in order:

(i) In the calculation of time dependent topological indexes, sudden jumps of topological indexes 
may be encountered in the long-time dynamics\cite{sacramento2014fate}. This artificial result arises from the fact that any two initial wave functions with slightly different momenta, after long time dynamics, can become totally different and even orthogonal. In this case, the approximations in the calculation of topological indexes
are no longer valid\cite{fukui2005chern,fukui2007quantum,fukui2007topological}. That means, in the long time dynamics, a much denser discretization of the parameter space is required to obtain the correct index. 
 Although the topological indexes are unchanged, signatures from the edge modes still allow us to explore the intriguing 
dynamical physics across the topological phase boundaries. For example, we may found that the zero-energy edge modes can oscillate periodically between the two edges 
in a clear Kitaev chain\cite{sacramento2014fate}, however, this phenomena is quite fragile, and the edge modes will quickly merge to the bulk bands up on weak disorder. 

(ii) This conclusion does not mean that it is impossible to transfer a topological state to a trivial state. Two possible strategies can be used to achieve this goal. Notice that 
all our analysis depends strongly on the unitary transformation and the auxiliary Hamiltonian with the same symmetries as the original Hamiltonian. In the first case, the unitary feature
can break down in the presence of dissipation\cite{hu2016dynamical, hu2015majorana}. In the latter case, topological phase transitions even in the adiabatic limit without gap closing can be
realized, if the time dependent wave functions violate some of the symmetries\cite{ezawa2013topological}. These two cases are beyond the scope of this work.

(iii) For the static Hamiltonian problem, the topological indexes can be generally computed using Green's function approach, which is more widely used in literatures\cite{wang2012simplified, 
wang2010topological}. With the auxiliary Hamiltonian, it is possible to construct the instantaneous Green's function as $\mathcal{G} = \sum_n (i\omega - \varepsilon_{n}({\bm k}))^{-1} 
|\psi_{n}({\bm k}, t) \rangle \langle \psi_{n}({\bm k}, t)|$, with which it is possible to investigate the dynamics of topological indexes with many-body interactions\cite{wang2010topological}.

To conclude, we demonstrate that the topological indexes during symmetry preserving dynamics are independent of time. We understand this general result from different angles. Different from the previous literatures, here, this conclusion has generality and can find wide applications in investigating the quench dynamics of different topological models.

\textit{Acknowledgements.} We thank the support of the National Youth Thousand Talents Program (No. KJ2030000001), the USTC start-up funding (No. KY2030000053)
and the NSFC (No. GG2470000101), and the  ``Strategic Priority Research Program(B)'' of the Chinese Academy of Sciences, Grant No. XDB01030200. B.H. is supported by the 
Natural Science Foundation of Jiangsu Province under Grant No BK20130424, and the National Natural Science Foundation of China under Grant No 11547047.

\bibliography{ref}

\end{document}